# DEVELOPMENT OF A DATA-DRIVEN TURBULENCE MODEL FOR 3D THERMAL STRATIFICATION SIMULATION DURING REACTOR TRANSIENTS

**Yangmo Zhu, Nam Dinh**
Department of Nuclear Engineering
North Carolina State University
2500 Stinson Dr, Raleigh, NC 27607
yangmo_zhu@ncsu.edu; ntdinh@ncsu.edu

**Rui Hu, Adam Kraus**
Nuclear Science and Engineering Division
Argonne National Laboratory
9700 Cass Avenue, Lemont, IL 60439
rhu@anl.gov; arkraus@anl.gov

**ABSTRACT**

SAM, a plant-level system analysis tool for advanced reactors (SFR, LFR, MSR/FHR) is under development at Argonne. As a modern system code, SAM aims to improve the predictions of 3D flows relevant to reactor safety during transient conditions. In order to fulfill this goal, one approach is to implement modeling of turbulent flow in SAM through establishing an embedded surrogate model for Reynolds stress/turbulence viscosity based on machine learning techniques. The proposed approach is based on an assumption that there exists a functional dependency relationship between local flow features and local turbulence viscosity or Reynolds stress. There have been very limited studies performed to validate this assumption. This paper documents a case study to examine the assumption in a scenario of potential reactor applications. The work doesn't aim to theoretically validate the assumption, but practically validate the assumption within the limited application domain.

From the methodological point of view, the approach used in this paper could be classified into the so-called Type I machine learning (ML) approach, where a scale separation assumption is proposed claiming that conservation equations and closure relations are scale separable, for which the turbulence models are local rather than global.

The CFD case studied in this work is a 3D transient thermal stratification tank flow problem performed using a Reynolds-averaged Navier-Stokes turbulence model in STARCCM+ code. Flow information of all geometric points in all timesteps is collected as training data and test data. The work aims to answer one question: is it possible to establish a surrogate model to model the functional dependency relationship between local flow features and local turbulence viscosity? For the case study, this paper practically validates the basic assumption in the methodological approach of implementing a data-driven surrogate model for turbulent viscosity into system analysis tool SAM. Future work will focus on the coupling between SAM and the surrogate model.

**KEYWORDS**
Turbulence, data-driven method, CFD, RANS

## 1. INTRODUCTION

A modern plant-level system analysis tool for advanced reactors (SFR, LFR, MSR/FHR) SAM is under development at Argonne National Laboratory, with advances in software environments and design, numerical methods, and physical models[1]. While focusing on system thermal fluid simulations, it has enhancements in the core and large volume modeling. It is also very flexible to be coupled with other physics or high-fidelity tools for multi-scale multi-physics simulations.

The flow modeling in large enclosures is a common phenomenon important to reactor safety in a nuclear power plant. In particular, flow mixing and thermal stratification in large volumes (e.g., lower plenum, upper plenum, downcomer) in both steady and abnormal transient scenarios are the the primary focus of this study. Two methods have been commonly applied in the simulation and analysis of thermalfluids in such scenarios. The first method is the Reynolds-averaged Navier-Stokes (RANS) -based CFD simulation, which has the following features: (i). RANS simulation is multi-dimensional; (ii). Turbulence models (including wall functions) are needed; (iii). CFD simulation is usually computationally intensive compared to system-level code; (iv). Various sources of uncertainties exist in CFD, including numerical errors. The other method is the system-level thermal-hydraulic code, which has the following features: (i). System code is usually 0-D model or 1-D model; (ii). It uses empirical correlations for special phenomena such as jet or plume of paramount importance for scenarios under consideration; (iii). Large uncertainties are contained in this method, especially considering the complexity in jet behavior affected by geometry and buoyancy effects. In short, those two methods both have advantages and disadvantages. Potentially, an advanced capability for plant-level simulation should leverage the advantages of both of the methods. Coupled system code and CFD code simulations have also been applied. Traditionally, CFD simulation and system-level simulation are performed separately with different software, which has its own limitations associated with code coupling.

As a new next-generation system simulation tool, SAM is designed to incorporate both multi-dimensional flow and system-level simulation simultaneously. The initial multi-dimensional flow simulation capabilities has been demonstrated in a previous work [2]. In order to fulfill the final goal: the integrated multi-dimensional and system-level flow simulation in a single SAM simulation, separated research topics have been established. One of them is the simplification of turbulence models and wall functions in 3D flow simulations: a highly parameterized and integrated turbulence model is expected.

Traditionally, the approach of enabling RANS-based CFD code to solve turbulent flow problem is to add turbulence models and wall functions to the code, which are not only various in amount and category but also flow pattern-dependent and hence scenario sensitive. Recent years, in view of the development of machine learning (ML) in turbulence modeling area [3-8], a trend emerged that uses a large amount of high-fidelity CFD-generated data to train a surrogate ML model to fulfill the role of turbulence models in CFD code.

According to Chang's classification [9], the framework of using machine learning to improve thermal-fluid modeling could be classified into 5 types. The approach used in this paper is so-called Type I ML, where a scale separation assumption is proposed claiming that conservation equations and closure relations are scale separable, for which the models are local rather than global. Several studies have already been performed in this area [3,4,9]. Ma [4] leveraged the approach to solve simple bubbly system problem; Duraisamy [3] selected global data as model input to solve turbulence flow problem. (In difference from [3], the present work selects local data as model input in order to achieve stronger portability and compatibility to the trained model.) Chang and Dinh [9] further applied the approach to a 2D steady-state flow problem. Now, 3D transient flows are studied with thermal stratification phenomena in this paper.

The choice of machine learning to train turbulence model in SAM is rooted in four potential advantages of ML surrogate model as compared to the traditional turbulence models: first, the ML approach could increase efficiency in the model (and code) development, by utilizing computational power; second, ML approach has the potential to capture behavior not limited to known "mechanisms" or models; third, potentially reduce research biases and implementation errors; and fourth, the ML surrogate model satisfies the SAM required modeling structure: highly parameterized and integrated.

The turbulence flow analyzed in this work's case study is a 3D thermal stratification (data-driven) problem. Data of each geometric points and timesteps are used to train the surrogate model, with total data amount of 3TB.

## 2. FRAMEWORK

This section outlines the framework of data-driven turbulence modeling. Its key procedures and components are discussed.

The framework of how data-driven turbulence modeling serves as a part of SAM is shown in Figure 1:

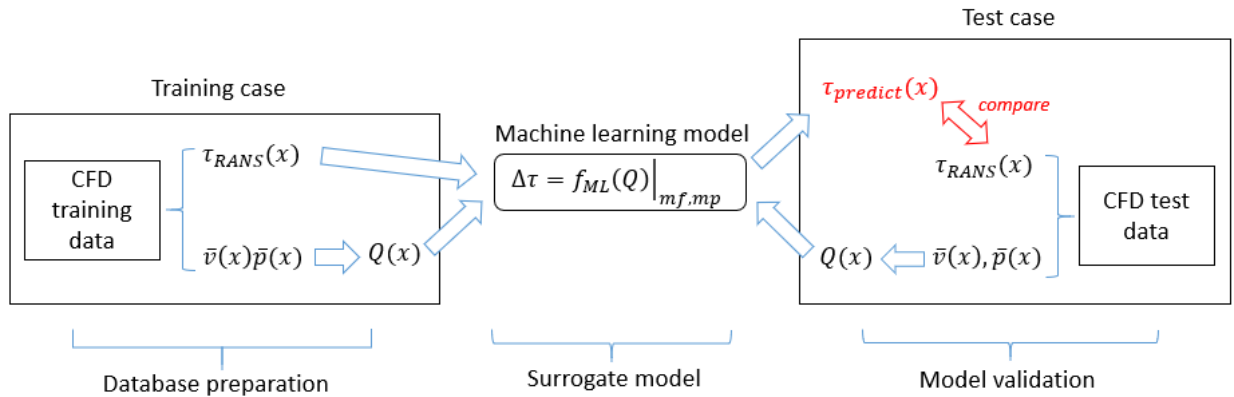

**Figure 1. The framework of the data-driven turbulence modeling problem**

The overall procedure can be summarized as follow:
1. Database preparation. In this step, turbulence properties (Reynolds stresses) $\tau$ are extracted from CFD training data, which are considered as output response of the machine learning model. Flow features $Q$ are obtained from basic flow quantities extracted from CFD results, which are considered as input for the machine learning model;
2. Surrogate model development. In this step, a data-driven regression function $f_{ML}$ is trained between flow feature $Q$ and turbulence properties $\tau$ with machine learning algorithm;
3. Model validation. In this step, the trained data-driven regression function is taken to predict Reynolds stresses $\tau$ on CFD test data. Then, the predicted Reynolds stresses are compared with the true Reynolds stresses of test data to validate the surrogate model.

### 2.1. Input and response of the machine learning model

The structure of the ML surrogate model could be seen as follow:

Because the role of the ML surrogate model is to replace the position of traditional turbulence model, the definition of the ML surrogate model is defined as:

$$\tau = f_{ML}(\mu, p, T, \nabla T, u, \nabla u, d) \tag{1}$$

Here we make the first assumption that there is a 1 vs 1 mapping relationship between the Reynolds stresses and a set of model inputs, which include:
- Dynamic viscosity (Pa-s)
- Pressure (Pa)
- Temperature (K)
- Temperature gradient (K/m)
- Velocity (m/s)
- Velocity gradient ($s^{-1}$)
- $d$, distance to the nearest wall (m)

The ML model target is Reynolds stress. Further, according to Boussinesq eddy viscosity assumption:

$$\tau_{ij} = 2\mu_t S_{ij}^* - \frac{2}{3}\rho k \delta_{ij} \tag{2}$$

Where $\mu_t$ is a scalar property call the turbulence viscosity. $S_{ij}^*$ is the mean strain rate tensor, $\rho$ is the fluid density, $k$ is the kinetic energy, and $\delta_{ij}$ is the Kronecker delta function. Because $S_{ij}^*$ is a function of velocity gradient, and fluid density is a function of temperature, Equation (1) could be further simplified into:

$$\mu_t = f_{ML}(\mu, p, T, \nabla T, u, \nabla u, d, k) \tag{3}$$

Here we make the second assumption that the effect of kinetic energy $k$ to turbulence viscosity $\mu_t$ is small enough to be neglected:

$$\mu_t = f_{ML}(\mu, p, T, \nabla T, u, \nabla u, d) \tag{4}$$

The reasons we make the second assumption are:
- Technically, such treatment could reduce the number of surrogate models from 6 to 1, which would largely increase the computational efficiency and robustness of the code.
- Theoretically, the kinetic energy $k$ couldn't be obtained in Type I ML approach.

### 2.2. The workflow in training the ML surrogate model

In this study, we use Deep Feedforward Neural Network (DFNN) technique to construct the surrogate model [10]. The neural network has the following advantages: (i) The method is well tested and implemented in widely used software packages, e.g., TensorFlow, Pytorch, etc; (ii) No matter how much time used in the training part, the test part is very fast. But it also has the disadvantage of the complexity in optimizing hyperparameters. It not only depends on experience but also depends on the specific case.

The workflow of training the surrogate model could be seen in Figure 2:

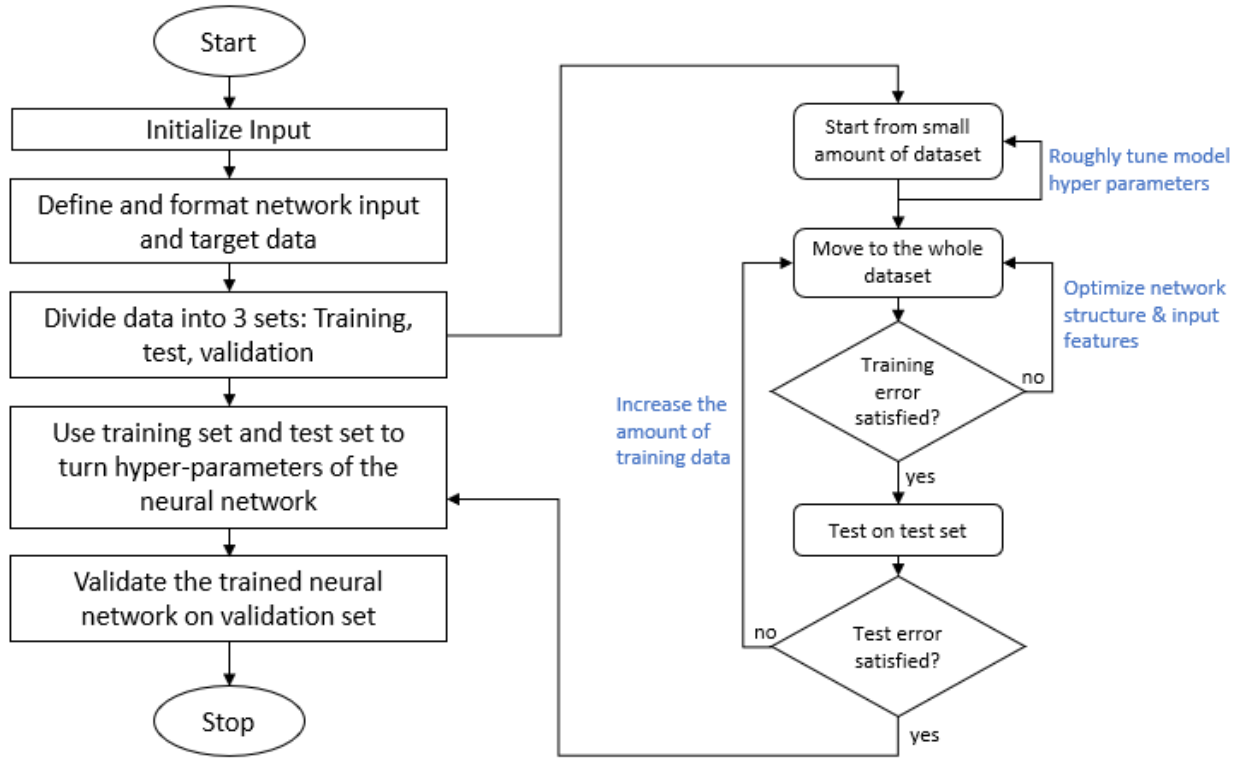

Figure 2.  The workflow: how to "babysit" your model

The workflow starts from initializing input, followed by training, test and validating the model. The major parts are dividing data into training, test and validation sets, and use the training set and test set to optimize hyper-parameters of the neural network.

Firstly, the total data amount is very large, 3TB (this large amount is because we collected data of each grid point in each time step). In order to deal with such large data, the model should be trained first on a small dataset to roughly optimize the hyperparameters in the neural network.

After major hyper-parameters are set, the training data could then gradually be extended to a bigger dataset. During this process, training and test should be done iteratively to finally obtain an appropriate setting. If the training error is higher than acceptable stop criteria, more complex network is recommended to be applied.

If the training error is less than acceptable stop criteria but the test error is higher than acceptable stop criteria, more training data is recommended to be added into the dataset.

In a general case, a detailed experiment record table should be formulated to help optimizing hyperparameters. The reason is usually because of the highly unpredictable trend in optimizing hyperparameters.

The present study considered application of batch normalization technique, which is a newly developed technique to achieve faster convergence. Upon evaluation for the case at hand, it is observed that batch normalization does not reduce computational expense. The convergence of training error becomes better with batch normalization, but each batch requires longer training time, hence then total training time becomes longer. Consequently, in the case study, batch noaalization was not applied in the network.

## 3. CASE STUDY: 3D THERMAL STRATIFICATION

### 3.1. Case Setup

The case studied in this research is a 3D thermal stratification problem.

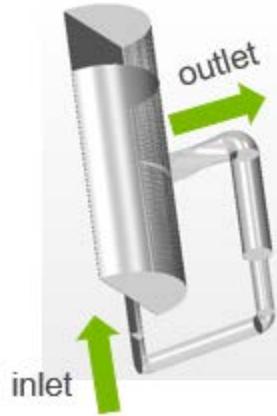

**Figure 3. An illustrative picture of the case geometry**

As can be seen in Figure 3, the fluid flows into the bottom pipe and flows out through the middle pipe. The tank is 8m in height and 2m in radius. Inlet pipe radius is 0.3m, and output pipe is an ellipse pipe with 0.2m as the semi-minor axis and 1m as the semi-major axis.

The case aims to simulate a transient scenario where the inlet flow rate gradually stops to zero, inlet flow temperature decreases at the first few seconds, then gradually recover to the initial value. The trend of flow rate and temperature could be seen in Figure 4.

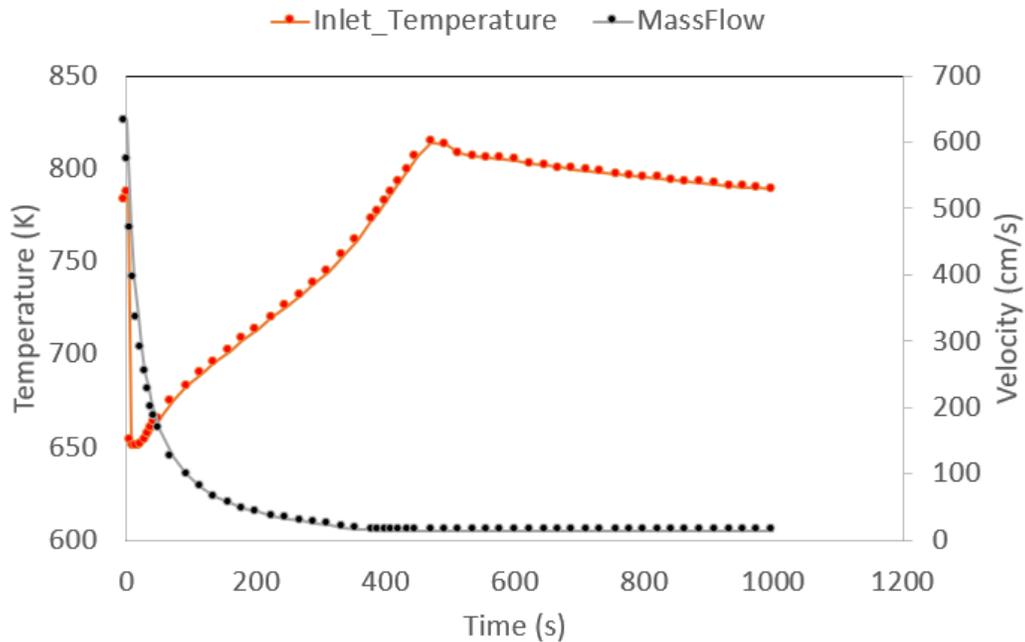

**Figure 4. Inlet boundary condition with time**

### 3.1.1. Step 1: roughly optimize hyperparameters on a small training dataset

In dealing with a large amount of data, it is unavoidable to face the problem of "imbalanced data". The term referring to the situation that data are not evenly distributed in all input domain. Data are redundant in some input domain while sparse in other input domain. The redundant data would largely increase the computational expense in optimizing the hyperparameters. Usually, resampling techniques, including over-sampling and under-sampling, is applied to balance the data [11].

It is noted that because of the limitation in project time and computational resources, the present study does not include a data resampling process. Such a choice is rationalized by the following arguments:
- The significant work time to develop a resampling algorithm for our case study;
- Resampling algorithm is not a major goal of this study;
- Neglect resampling has a minimal impact on the final prediction accuracy of the ML surrogate model.

In this work, we apply a simpler way to deal with the "imbalanced data" problem. The procedure is described shortly below, and needs to be viewed as a tentative solution. As moving forward, we recommend to develop a resampling algorithm, which would save computational expense in the long run while minimizing information loss.

Because of the "imbalanced data" problem, training time is too long to be accepted. The way we selected to deal with this problem is to "arbitrarily delete" training data amount. Different from under-sampling approach resampling technique, which only deletes redundant information, the term "arbitrarily delete" defined here is deleting data according to the researcher's experience and intuition. Consequently, so useful information may also be removed.

In the process of arbitrarily deleting data, the authors follow the following principles:

- The whole data comes from one simulation of a long transient scenario. The scenario starts from transient, and gradually approach a steady state. It is assumed that the more dramatic the transient is, the more useful data it provides. Data in these timesteps should be kept more than those in steady state.
- Data are only deleted in the time domain, not in the geometric domain. For example, if we have the simulation data of a 10x10 mesh flow in 20 timesteps, we will delete data in the time domain, say only keep 2 timestep's data, so the final data amount is 10x10x2.

After using "arbitrarily deleting" process to compress data, the remain data amount (4.2% of the original data) is still too large for roughly optimize major hyperparameters. Here we use a 2-step strategy to optimize the hyperparameters:

1. Only use 0.1% of the total data from the 4.2% remain data as training/validation/test data to roughly determine major hyperparameters;
2. Keep the major hyperparameters fixed, extend the data from 0.1% to 4.2% to future optimization of other hyperparameters.

A way of selecting the 0.1% data as recommended by machine learning experts is a random selection, so to reduce human bias. But the way the authors choose is selecting the 0.1% data on a certain small-time range. In this case, the total time steps are 5591, we select the first 450 timesteps data, and extract part of these data as training data, test the model on other data in the first 450 timesteps. The advantage is to quickly test whether the model could make a good prediction in a short time range or not. If it couldn't even make a prediction on small time steps, it suggests error in the model's other components, for example, selection of model input & target, network structure, and optimization algorithm.

The final optimized model hyperparameters are as following:

1. Network structure: 5-layer network. 4 ReLU layer + Tanh layer
2. Number of nodes in each layer: 1000
3. Batch normalization is turned off as they extremely slow down training speed (batch normalization could make train converge in fewer iterations, but each iteration would cost more time to train, so overall it will delay training speed. If the computational resource is enough this option is suggested to be turned on.)
4. Learning rate: 0.0001
5. Learning rate decay: decay 0.3 after 200 iterations
6. Batch size: 2^18 = 262144
7. Number of workers = 4
8. Stopping Criteria: training error less than 0.00003
9. Optimization algorithm: ADAM (adaptive moment estimation) algorithm

### 3.1.2. Step 2: extend the training dataset

When moving to a large dataset, the process follows the flowchart (Figure. 2) to gradually increase the data amount. Because the whole data is too large to be used as training data, it is important to find the minimal set of training data when dealing with large data.

Currently, 4.2% of the total data are used as training data, such training could cost 2 weeks to reach 3e-5 training error with 1 node of GPU on BLUES (cluster as Argonne national lab). Parallel training is also tested but the performance is not optimized (only 5% faster). Consequently, only 1 GPU node is used to train the model. Given more time and computational resources, model data could be used, and better accuracy is expected to be achieved.

### 3.2. Numerical Results

### 3.2.1. Training data, validation data and test data

The training data and validation data are local data of each point in certain timesteps. The timesteps are selected as below:

In the first 10 timesteps, all data are used. After that, we sample data every 20 timesteps. After 450 timesteps, we sample data every 25 timesteps until 5500 timesteps, which approaches the steady state.

In order to test the model, the test dataset is composed. The test dataset is composed based on the "worst situation principle", which means use the data that potentially has the worst prediction result, so to test what the model would perform in the worst situation. For example, if training data is in timestep 50 and 70, then the data in timestep 60 is considered as potentially would have the worst performance, and thus would be selected as test data.

### 3.2.2. Performance: prediction accuracy

The surrogate model is evaluated by analyzing the prediction accuracy. The prediction accuracy is defined here as the portion of test data that satisfies the required test error divided by the total amount of test data.

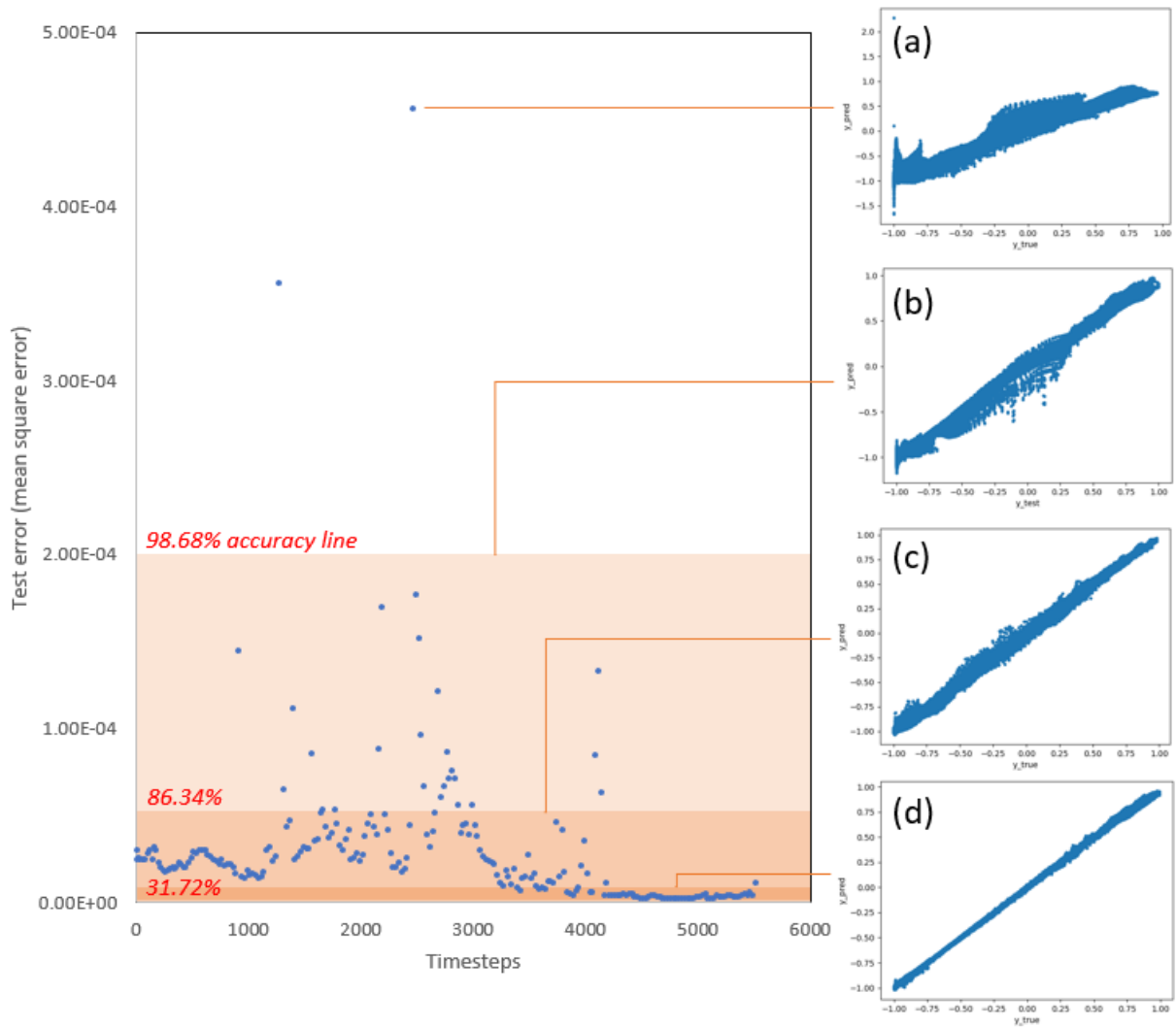

**Figure 5. Prediction accuracy for the different acceptable error range**

As can be seen in Figure 5, each point in the left picture represents the mean square error of all the data in that time step. On the right-hand side, there are 4 pictures, which illustrate how the number of test error correlated with actual performance between true turbulence viscosity and model predicted turbulence viscosity. For Figure 5 (a,b,c,d), the x-axis is true turbulence viscosity and the y-axis is model predicted turbulence viscosity. As can be seen, the worst performance could be like Figure 5(a); if the test error is equal to 2e-4, then the actual correlation could be like Figure 5(b), which is better than Figure 5(a); if the test error is equal to 5e-5, then the actual correlation could be like Figure 5(c), which is better than Figure 5(b); if the test error is equal to 1.16e-5, then the actual correlation could be like Figure 5(d), which is better than all of the above correlations. So, the smaller the test error, the better the model.

The acceptable error range is another important term. It is defined here as the acceptable error that could still allow CFD calculation to perform without diverging or running into a wrong solution. If the acceptable error range is 2e-4, then nearly all the points are below this value, the prediction accuracy is 98.68%. If a better model is required, then the acceptable error range should be reduced accordingly. When acceptable error range is equal to 5e-5, the prediction accuracy is 86.34%; when the acceptable error range is equal to 1.16e-5, the prediction accuracy is 31.72%.

It could be seen from Figure 5 that, after 4000 timesteps, the prediction result becomes much better. This may be related to the fact that the flow becomes increasingly steady. This result indicates that our model performs better for steady state flow than for transient flow.

Time derivative terms may need to be added into the model input/flow feature. In the first 450 iterations, we use 5% data as training data; In the following 5000 iterations, we use 4% data as training data. So, the data sample rate is very similar, but the performance differs noticeably, especially in 1000~4200 timesteps, which is the time flow changes dramatically due to thermal stratification.

A similar study has also been performed with a reduced amount of training data to investigate the degree by which the training data amount would affect the final prediction accuracy.

**Table I. Prediction accuracy of models trained by different amount of training data**

| Acceptable error range | Prediction accuracy (use 2.4% data as training data) | Prediction accuracy (use 4.2% data as training data) |
|---|---|---|
| 2e-04 | 91.06% | 98.68% |
| 5e-05 | 59.54% | 86.34% |
| 1.16e-05 | 21.14% | 31.72% |

As can be seen in Table 1, increasing training data amount from 2.4% to 4.2% of total data significantly improve the model prediction accuracy. It is believed that with more data being used as training data, higher prediction accuracy could be achieved.

It is noted that for the work presented in this section we only perform analysis of the result for test data. Result for validation data is not shown as it is not important: The model hyperparameters are optimized until validation data achieve the same prediction error and training data. So it would be redundant to show validation data performance.

## 4. DISCUSSION AND FUTURE WORK

A 3D transient thermal stratification tank flow problem has been selected to perform to examine the fundamental assumption for Type I ML approach (as detailed in section 2): is it possible to establish a surrogate model to represent the functional dependency relationship between local flow features and local Reynolds stresses/turbulence viscosity?

According to the result, in the common application domain of SAM (represented by a 3D thermal stratification scenario), it is very likely that a functional dependency correlation between local flow features and local Reynolds stress could be established. We use the term "very likely" because the final "yes" or "no" answer would depend on the acceptable error range and prediction accuracy required in the analysis.

There are areas for further improvement to achieve better model performance and higher work efficiency:
- For data generation part: to increase model accuracy, sufficient computational resource should be allocated, while the current computational resource is not enough to support 3D simulations. One solution is to switch to 2D cases to reduce computational cost; the other solution is to use a "transfer learning" technique to improve the current model so that to save time. In any caase, the computational expense for training of the ML model is still an important constraint;
- For data pre-processing part: Resampling algorithm is suggested to be selected and adapted, which could improve work efficiency;

- For feature engineering part: time derivative terms are suggested to be added into local flow features, which may improve model performance.

After increasing model accuracy to an acceptable level, the next step would be integrating the model into SAM, which would bring additional challenges. Prediction error in turbulence viscosity would be amplified after several iterations. The possible solution to this problem is to train the model with more training data from different cases.

## 5. CONCLUSIONS

In this study, we study the feasibility of implementing a data-driven surrogate model for turbulent viscosity for thermal mixing and stratification modeling into the system analysis tool SAM. According to the case study result, the trained surrogate model shows 98.68% prediction accuracy if the acceptable error range is set to be 2e-4. Also, the result shows the trends of prediction accuracy increasing with data amount. This suggests that a functional dependent correlation between local quantities of interest and local Reynolds stress is very likely to exist.

The methodological, technical and practical values of this study can be summarized as below:
- In the methodological point of view, this study explores the potential use of high-fidelity RANS data rather than DNS. When justified, it could largely save computational expense for obtaining the training data.
- In the technical point of view, this study documents the workflow and the experience in dealing with a large amount of data, which is inevitable in Type I ML.
- In the practical point of view, this study validates the fundamental assumption of Type I ML, which is instrumental to account for the turbulence effects in SAM 3D simulations.

## ACKNOWLEDGMENTS

This work is supported by U.S. DOE Office of Nuclear Energy's Nuclear Energy Advanced Modeling and Simulation (NEAMS) program. We gratefully acknowledge usage of the Blues and Bebop clusters in the Laboratory Computing Resource Center and the Eddy cluster in the Nuclear Science and Engineering Division at Argonne National Laboratory. A partial support FOR Y.Z and N.D. from the U.S. Department of Energy's Consolidated Innovative Nuclear Research program via the Integrated Research Project on under the grant DE-NE0008530 is acknowledged.